\documentclass[a4,twocolumn]{aa}

\usepackage{graphicx,txfonts,natbib,verbatim}
\bibpunct{(}{)}{;}{a}{}{,}

\begin{document}

\title{Multi-thermal dynamics and energetics \\ of a coronal mass
ejection in the low solar atmosphere}

\author{I. G. Hannah \and E. P. Kontar}

\offprints{Hannah \email{iain.hannah@glasgow.ac.uk}}

\institute{SUPA School of Physics \& Astronomy, University of Glasgow, Glasgow,
G12 8QQ, UK}

\date{Received ; Accepted }

\abstract{}{The aim of this work is to determine the multi-thermal
characteristics and plasma energetics of an eruptive plasmoid and occulted flare
observed by Solar Dynamics Observatory/Atmospheric Imaging Assembly
(SDO/AIA).}{We study an event from 03-Nov-2010 (peaking at 12:20UT in GOES soft
X-rays) of a coronal mass ejection and occulted flare which demonstrates the
morphology of a classic erupting flux rope. The high spatial, and time
resolution, and six coronal channels, of the SDO/AIA images allows the dynamics
of the multi-thermal emission during the initial phases of eruption to be
studied in detail. The Differential Emission Measure (DEM) is calculated, using
an optimised version of a regularized inversion method (Hannah \& Kontar 2012),
for each pixel across the six channels at different times, resulting in emission
measure maps and movies in a variety of temperature ranges. }{We find that the
core of the erupting plasmoid is hot (8-11, 11-14MK) with a similarly hot
filamentary ``stem'' structure connecting it to the lower atmosphere, which
could be interpreted as the current sheet in the flux rope model, though is
wider than these models suggest. The velocity of the leading edge of the
eruption is 597-664 km s$^{-1}$ in the temperature range $\ge$3-4MK and between
1029-1246 km s$^{-1}$ for $\le$2-3MK. We estimate the density (in 11-14 MK) of
the erupting core and stem during the impulsive phase to be   about
$3\times10^9$ cm$^{-3}$, $6\times10^9$ cm$^{-3}$, $9\times10^8$ cm$^{-3}$ in the
plasmoid core, stem and surrounding envelope of material. This gives thermal
energy estimates of $5\times10^{29}$ erg, $1\times10^{29}$ erg and
$2\times10^{30}$ erg. The kinetic energy for the core and envelope is slightly
smaller. The thermal energy of the core and current sheet grows during the
eruption, suggesting continuous influx of energy presumably via
reconnection.}{The combination of the optimised regularized inversion method and
SDO/AIA data allows the multi-thermal characteristics (i.e.
velocity, density and thermal energies) of the CME eruption to be determined.}

\titlerunning{Multi-thermal dynamics and energetics of a coronal mass
ejection}

\authorrunning{Hannah \& Kontar}
\maketitle

\section{Introduction}

The solar atmosphere is awash with highly dynamic phenomena driven by the rapid
liberation of magnetically stored energy. Magnetic reconnection is thought to
facilitate this energy release, resulting in solar flares, coronal mass
ejections (CMEs) and the associated particle acceleration, plasma heating and
bulk motions. The classic model of solar eruptive events (SEE) is the CSHKP
model
\citep{1964NASSP..50..451C,1966Natur.211..695S,1974SoPh...34..323H,1976SoPh...50...85K}
which continues to become more ornate over time \citep[for
instance][]{2000JGR...105.2375L}. The basic scenario is that of an erupting flux
rope (helical magnetic structure) which stretches and elongates the coronal
magnetic field behind it, instigating magnetic reconnection in a vertical current sheet. This
releases considerable energy, particularly in accelerated particles that then
travel down into the denser lower solar atmosphere. Here they lose their energy
through collisions with the background plasma, producing hard X-ray (HXR)
footpoints and flare ribbons, heating material that then expands upwards filling
flaring loops.

Flares are well observed in the lower solar atmosphere, but CMEs have, till
recently, mostly been studied at higher altitudes (for an overview see
\citet{2006SSRv..123...13H,2012LRSP....9....3W}) due to the lack of suitable
high cadence observations. This means that there is considerable uncertainty as
to the specific processes that develop flux ropes and initiate CMEs
\citep[e.g][]{2010ApJ...718.1388D}. High in the corona, white-light
observations typically show CMEs structured as a  bright loop ahead of a darker
cavity containing a bright core. Here the bright loop is taken to be due to the
increased density of material amassed in the leading front of the eruption with
the cavity likely due to to a flux rope
\citep{2006ApJ...641..590G}. CMEs have been found to evolve in distinct
stages, initiation, acceleration and propagation \citep{2006ApJ...649.1100Z} and
the initial acceleration stage of CME synchronize well with the flare HXR
\citep{1988ApJ...328..824K,2005ApJ...630.1148S,2008ApJ...673L..95T,
2010ApJ...712.1410T}
and soft X-ray (SXR) emission \citep{2001ApJ...559..452Z,2005A&A...435.1149V},
while the HXR flux correlates with the acceleration of CMEs
\citep{2010ApJ...712.1410T}.  Active regions are significantly more likely to be
eruptive if they have a sigmoidal morphologly \citep{1999GeoRL..26..627C} and
these structures are seen as the precursors to flux ropes both observational
\citep{2010ApJ...725L..84L,2011A&A...526A...2G} and in simulations
\citep{2003A&A...406.1043T,2003ApJ...589L.105F,2008A&A...492L..35A,
2010ApJ...708..314A}.

 \begin{figure*}\centering
\includegraphics[width=140mm]{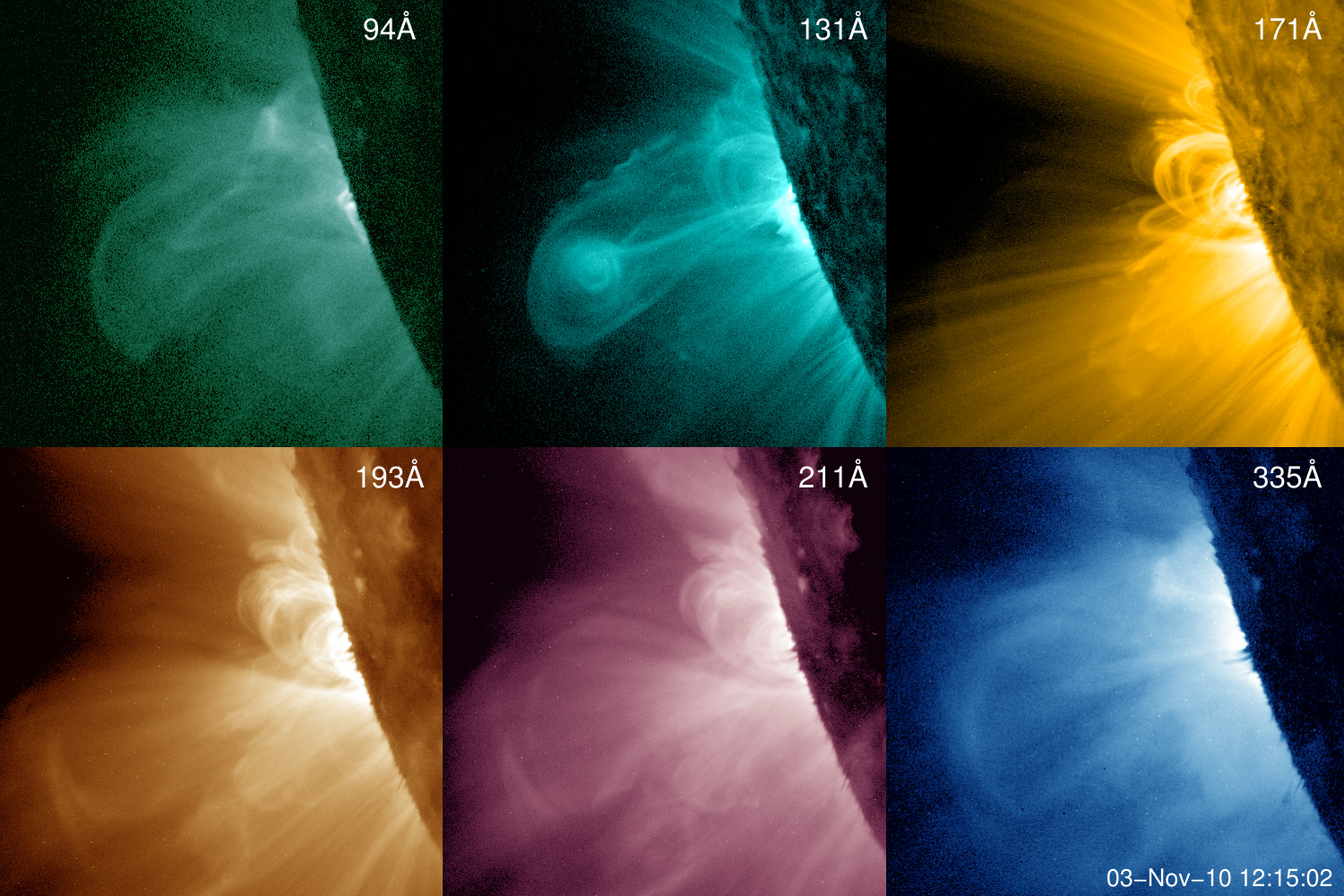}
\caption{\label{fig:aia121502}SDO/AIA observations from each of the six coronal
filters during the impulsive phase of the 3-Nov-2010 flare when the erupting
plasmoid is clearly visible in 131\AA. The time shown is for the 94\AA~ image,
with all from 12:15:00 to 12:15:09UT.}
\end{figure*} 

Associated with CMEs is the nearby dimming of the corona during the
onset stage, seen in soft X-rays \citep{1997ApJ...491L..55S} and EUV
\citep[e.g.][]{1998ApJ...498L.179G}.
This could be due to the depletion of material as part of the eruption, plasma
heating or a combination of both. Observations generally point to the former:
for instance five CMEs studied by \citet{2003A&A...400.1071H} using EUV
spectroscopic data from CDS showed dimming due to mass loss and not temperature
variations. However the are also examples  of CMEs where the dimming is due to
plasma heating and not depletion \citep{2010ApJ...720L..88R}.

Spectral observations of CMEs in the extended corona ($>1.5R_{\sun}$
from Sun centre) with SOHO/UVCS revealed many aspects of CMEs at these altitudes
\citep{2006A&ARv..13...31K}, observing spectral lines from CMEs over 0.02 -- 6
MK.  In one CME a ``narrow'' (100Mm) structure was observed for 20hrs, with an
emission measure of about $10^{25}$cm$^{-5}$ over  4 -- 6 MK connecting the CME
core to the flare loops below \citep{2002ApJ...575.1116C}.
This was interpreted as an edge-on view of the current sheet and a density of
$5\times10^{7}$cm$^{-3}$ was estimated using a 40Mm line-of-sight component.
Similarly hot current sheets have been observed in another long-duration event
\citep{2003ApJ...594.1068K} as well as faster ones \citep{2005ApJ...622.1251L}
where the leading edge and CME core were moving at 1\,939 and 1\,484 kms$^{-1}$
respectively. Hotter current sheets have been observed in some CMEs as well
\citep{2003SoPh..217..267I,2006ApJ...638.1110B,2010ApJ...708.1135S}.
\citet{2010ApJ...711...75L} used Hinode/EIS to determine the Differential
Emission Measure (DEM, see \S\ref{sec:demmeth}) of the CME lower in the solar
atmosphere (at $1.1R_{\sun}$). They found a multiple peaked DEM, with two
components  at 0.13 MK of density $10^{10-11}$ cm$^{-3}$ and another at 0.5 MK
of density $10^{8.7}$ cm$^{-3}$ travelling together.
The emission at about 1MK was due to the line-of-sight/background corona however
SXR emission was also detected by Hinode/XRT, enveloping the core and suggestive
of material at 5--10 MK. The analysis of the post eruption arcade found similar
plasma to the current sheet, both coronal plasma heated up to 3 MK
\citep{2012ApJ...751...21L}

High spatial ($\sim 1.5\arcsec$) and temporal resolution ($\sim 12$ s) EUV full
disk images in multiple wavelengths is now available from the Atmospheric Imaging
Assembly \citep{2012SoPh..275...17L}  on the Solar Dynamics Observatory
(SDO/AIA), with several authors taking advantage of this data to study CME
formation and propagation from the low corona
\citep[e.g.][]{2010ApJ...724L.188P,2011ApJ...733L..25K}.
One particular event, which peaked in GOES SXRs at 12:20UT 03-Nov-2010 (Figure
\ref{fig:aia121502}), of an occulted flare and erupting plasmoid, shows
unambiguously the formation of a flux rope and has subsequently been studied in
detail \citep{2011ApJ...727L..52R,2011ApJ...732L..25C,2011ApJ...729L...8F,
2012ApJ...750...44B,2012ApJ...754...13S,2012ApJ...761...62C}, and will be
studied in detail in this paper. This event shows the formation of a flux rope
clearly in 131\AA, images with a peak in their sensitivity at 11MK, though
dimming at the core in channels sensitive to cooler emission (171\AA, 211\AA),
shown in Figure \ref{fig:aia121502}. As the plasmoid move outwards, a
filamentary, stem-like structure forms behind it, again sharply defined in
131\AA, suggestive of a large-scale current sheet
\citep{2011ApJ...727L..52R,2011ApJ...732L..25C,2012ApJ...754...13S}.
The kinematics of the eruption in the different channels has been found to be
slower (about 600 kms$^{-1}$ versus 1\,400 kms$^{-1}$) in those sensitive to
hotter emission \citep{2011ApJ...732L..25C,2012ApJ...750...44B}.

The propagation of a metric Type II radio burst ahead of the CME in EUV is
faster still (nearing 2\,000 kms$^{-1}$) suggesting a piston-drive shock, in
which it is moving too quickly for the ambient plasma to flow behind the driver
\citep{2012ApJ...750...44B}. This work also found that HXR emission observed
with RHESSI had two non-thermal (18-40 keV) sources, one compact low in the
corona, the other extended and high in the corona, close to the erupting
plasmoid. The later could be due to electrons accelerated in the stem/current
sheet. Although this event superficially matches the cartoon model of flux rope
eruption things are more complicated:
\citet{2011ApJ...729L...8F} found Kelvin-Helmholtz waves along the top edge of
the erupting plasma, possibly allowing a secondary reconnection process.
\citet{2012ApJ...754...13S} studied the inflow and outflow into the CS region,
finding the hottest material inflowing fastest and correlating with the RHESSI
thermal X-ray emission. Pairs of inflowing and outflowing material was found to
be travelling away from the region faster than they arrived, suggesting 
acceleration.

Since SDO/AIA provides six different wavelength channels, sensitive to a
range of coronal temperatures, they can be used to determine the DEM of
the different parts of the CME. \citet{2012ApJ...761...62C} did this for this
event, as well as two others, producing DEMs for different CME regions: the
core of the erupting flux rope and the leading front.  They found the cores to
be have emission over a broad distribution of temperatures ($3\le T \le 20$MK),
with an average temperature of about $>8$MK and densities of about
$10^9$cm$^{-3}$. The leading front of the eruptions were cooler (about 2MK on
average) with a narrower temperature distribution ($1\le T \le 3$MK) similar to
the pre-eruption coronal material but with slightly higher densities. They
interpreted the leading fronts to be the showing signatures of compressed
ambient coronal plasma. Dimming, about 10~minutes after the eruption, was found
to be due to the rarefaction of material.

In this paper we use a different approach, regularized inversion, to
produce the DEM not for a few regions within the CME but for every single pixel
in the set of six SDO/AIA images,  every  12~seconds throughout the event. Thus
allowing the highly dynamic multi-thermal properties of the CME to be
investigated.  The regularized approach detailed in this paper
(see\S\ref{sec:demmeth}) is an optimized version of our original code,
introduced, tested, and compared to the approach used by
\citet{2012ApJ...761...62C} in \citet{2012A&A...539A.146H}. The version present
here is able to compute 10\,000s of DEMs per second, calculating both horizontal
and vertical and uncertainties (see \S\ref{sec:demregmet}) and without forcing a
prescribed DEM model. In \S\ref{sec:dempx} the properties of the DEM in
different pixels of the event are discussed, with the maps and their
uncertainties presented in \S\ref{sec:demmaps}. We then investigate the time
evolution of the DEMs in terms of the morphology, total emission and propagation
speed for the different temperature ranges, in \S\ref{sec:demtime}. As the
erupting plasmoid and filamentary stem structure is well defined at high
temperature (11-14MK), we use the emission measures in \S\ref{sec:deneng} to
estimate the densities and thermal energy of the different components.

\section{Differential Emission Measures}\label{sec:demmeth}
\subsection{Introduction}
The signal detected $g(x,y,t,i)$ in each pixel $(x,y)$ in the $i^{th}$ SDO/AIA
filter at time $t$ can be interpreted as

\begin{equation}\label{eq:dem_int}
g(x,y,t,i)=\int_T K(T,i) \xi(T,x,y,t) dT +\delta g(x,y,t,i),
\end{equation}

\noindent where $K(T,i)$ is the temperature response of the $i^{th}$ filter
\citep{2010A&A...521A..21O,2012SoPh..275...41B}.
The Differential Emission Measure  DEM $\xi(T,x,y,t)$, at a particular instance
in one pixel, is taken to be $\xi(T)=n(h)^2\mathrm{d}h/\mathrm{d}T$
[cm$^{-5}$K$^{-1}$], which is the electron density $n^2$ along the line-of-sight
$h$ of the emitting material at temperature $T$. The inherent noise in the
observations $\delta g(x,y,t,i)$, due to counting statistics, background and
instrumental uncertainties, will be significantly amplified when a direct
solution of Eq. (\ref{eq:dem_int}) is attempted.
This recovery of the DEM is an ill-posed inverse problem
\citep{ti63,1985InvPr...1..301B,1986ipag.book.....C,2006SoPh..237...61P} with
the growth of uncertainties resulting in spurious solutions. To recover
useful solutions of the problem given in Eq.
(\ref{eq:dem_int}), it needs to be constrained to avoid the noise amplification
in the DEM solution. The constraints are often physically motivated additional
information {\it a priori} known about the solution. An additional challenge
here is that we need an approach that can quickly and robustly calculate DEMs
since there is a wealth of SDO/AIA data to study this highly dynamic eruption.
The simple isothermal/ratio approach \citep[e.g.][]{1973SoPh...32...81V} places
a very strict, though possibly erroneous, constraint on the DEM solution and so
is ill-suited to this problem. An alternative is to forward fit a chosen DEM
model to the data, for example with multi-Gaussians
\citep[e.g.][]{2011ApJ...732...81A} or splines
\citep[e.g.][]{2004IAUS..223..321W}, the latter used for this CME event by
\citet{2012ApJ...761...62C}. Again, if the assumption about the form of the DEM
is wrong then the results are difficult to reliably interpret. These techniques
can be computationally slow, only allowing select regions to be analysed, not
taking full advantage of the SDO/AIA data. Other approaches, such as MCMC
\citep{1998ApJ...503..450K}, can produce very robust results, without requiring
predefined functional forms to the DEM solution, and uncertainty estimates but
are computationally slow and so are again inappropriate to obtain DEM maps.

Tikhonov regularized inversion recovers the solution by limiting the
amplification of the uncertainties through assumptions about its ``smoothness''
by using linear constraints
\citep{ti63,1977A&A....61..575C,2006SoPh..237...61P}. This approach does not
make any assumption on the functional form of DEM and we have implemented an
algorithm \citep{2012A&A...539A.146H} to recover the DEM from solar data using
Tikhonov regularization and Generalized Singular Value Decomposition GSVD
\citep{1992InvPr...8..849H}. This code, originally developed and extensively
tested for the inversion of X-ray data
\citep{2004SoPh..225..293K,2005SoPh..226..317K}, naturally provides estimates to
the uncertainties in the solution, an important feature, when trying to recover
a large number of DEMs. Recently, a simpler and computationally fast (computing
100s to 1,000s DEMs per second) SVD regularisation approach has been applied to
AIA data \citep{2012arXiv1204.6306P}, however it does not produce error
estimates. Regularization has also been used with solar data to determine the 3D
emission measure  through the use of solar rotation or multi-spacecraft
observations \citep{2005ApJ...628.1070F,2009ApJ...701..547F}. In general,
regularized approaches do not require a positive solution but one can be
obtained by using an iterative method such a Reiterated Conjugate Gradients,
which was used for solar data by \citet{1986SoPh..105..323F}. Our approach
uses one iteration to help with the weighting, and positivity, of the final solution
whilst maintaining the fast computational speed.

\subsection{Regularization Method for DEM maps}\label{sec:demregmet}

The full details of the method are in \citet{2012A&A...539A.146H} and this
previous version of the algorithm allows a variety of different controls to the
nature of the regularization but is too computationally slow to practically produce DEM
maps and movies (running at about a few DEMs per second). We have produced an
optimized and simplified version of the software\footnote{Available online:
http://www.astro.gla.ac.uk/$\sim$iain/demreg/map/}, detailed below with
considerable speed gains.  As the DEM calculation per pixel location is
independent of other pixels in the same image, further speed gains were achieved
by parallization of the code within IDL using Bridges. The results is that we
can achieve about 1\,000s DEMs per second on a standard multi-core cpu
laptop/desktop and 10\,000s on multi-cpu machines.

To obtain the DEM maps we perform regularization with the zeroth-order
constraint and no initial guess solution. For the detailed explanation of these
parameters and the affect they have on the regularized solution see
\citet{2012A&A...539A.146H}. The specific problem to be solved here is

\begin{equation}\label{eq:regdem}
\left\|{\bf \widetilde{K}}\;{\xi-\bf \widetilde{g}}\right\|^2
+\lambda\|{{{\bf L} \;\xi}}\|^2 =\mbox{min},
\end{equation}
\noindent where the constraint matrix ${\bf L}$ is the identity matrix
(normalised by the temperature bin width) and the tilde
indicates normalisation by the
error in the data $\delta g$. The GSVD singular values ($\gamma_i$, $\beta_i$)
and vectors ($\bf u, v, w$) produced from the GSVD of $\bf \widetilde{K}$ and $\bf L$ 
provide a ready DEM solution
$\xi_\lambda$ as a function of the regularisation parameter $\lambda$
\citep{1992InvPr...8..849H}, i.e.

\begin{equation}\label{eq:sol_gen}
\xi_\lambda(T) =\sum_{i=1}^ 6\frac{\phi_i^2}{\phi_i^2+\lambda}\left(
\frac{({\bf \widetilde g}\cdot {\bf u}_i){\bf w}_i}{\gamma_i}\right)\;,
\end{equation}

\noindent where $\phi_i=\gamma_i/\beta_i$. To find the regularization parameter
$\lambda$ that produces a solution $\xi_\lambda$ that matches the required
$\chi^2=1$ (in data space) we use coarser sampling of $\lambda$ than in the
original software which helps with the computational speed increase. Even with
this approach we achieve $\chi^2$ of typically between 0.5 and 1.75 with our
regularized solutions, which are identical (within the error bars) to the
results from the previous code.
The above procedure is run twice (as in the original code) with the result from
the first run $\xi_\lambda(T)$ being used to weight the constraint matrix as
$\mathbf{L}^2_{jj}=d\log{T_j}/\xi_\lambda(T_j)$ before the second GSVD is
performed and final $\xi_\lambda$ solution found. The horizontal errors are
again taken to be the spread of the resolution matrix $\mathbf{R_\lambda K}$ for
each temperature bin \citep{2012A&A...539A.146H}, where $\mathbf{R_\lambda}$ is
the regularized inverse assumed to be close to the generalized inverse
$\mathbf{K^\dagger}$, i.e. $\xi_\lambda\simeq\mathbf{R_\lambda g}$. For the
vertical error we do not use the Monte Carlo approach as in
\citet{2012A&A...539A.146H} but take it to be $\Delta \xi(T)^2=<\delta
\xi(T)^2>=\left\| \mathbf{R_\lambda \delta g} \right \|^2$. This produces a
similar result  but is considerably quicker to compute.

\begin{figure*}\centering
\includegraphics[width=150mm]{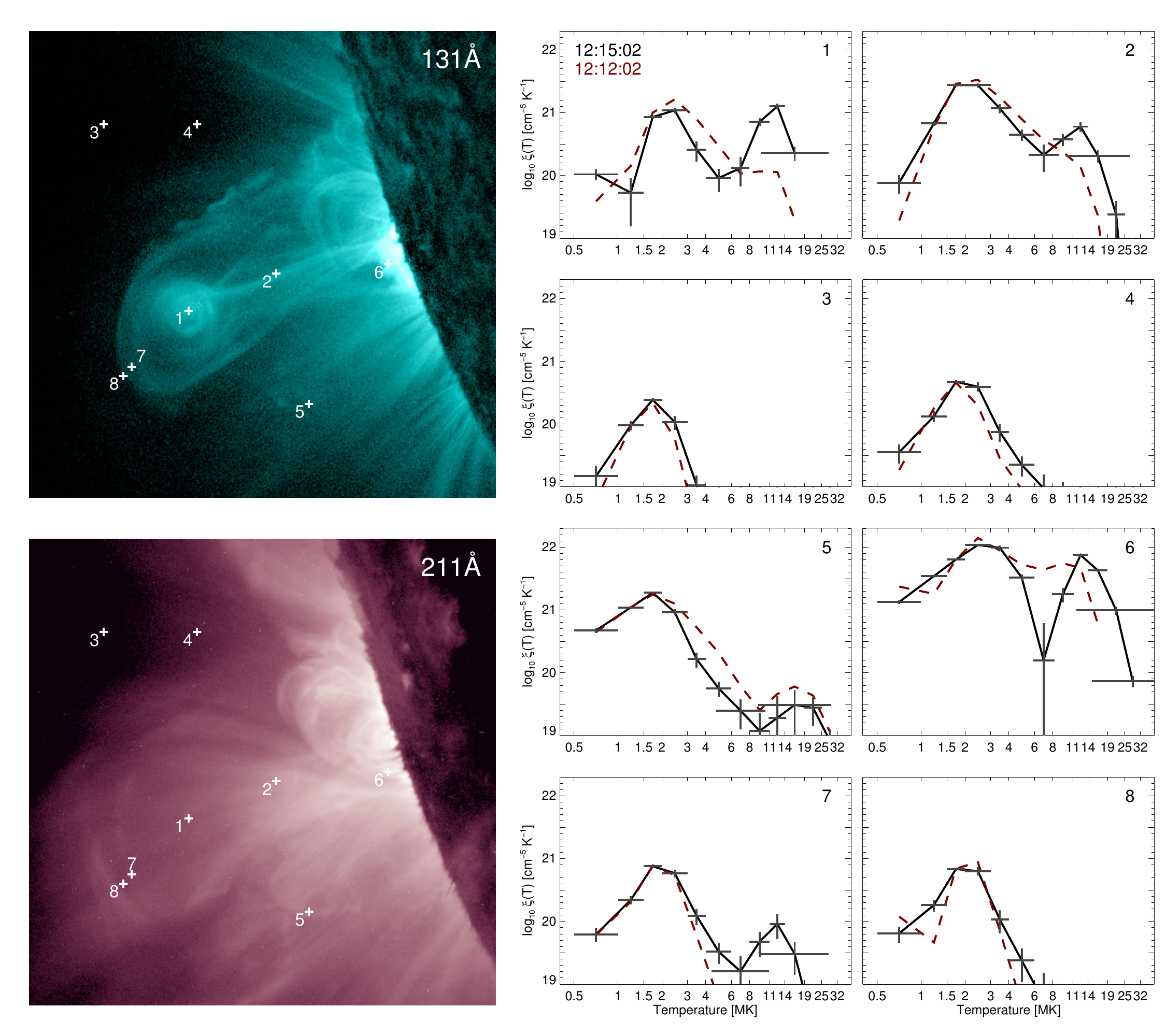}
\caption{\label{fig:DR121502}(Left) The 131\AA~and 211\AA~images from Figure
\ref{fig:aia121502} with eight pixels highlighted. (Right) The corresponding
regularized DEM solutions $\xi(T)$ (using twelve temperature bins) and the
vertical $\Delta \xi(T)$ and horizontal $\Delta \log{T}$ errors (the maximum
of the error from the resolution matrix or the temperature bin width) for the
highlighted pixels using the data from all six image in Figure
\ref{fig:aia121502}. The dashed lines show the DEM solutions in those
pixels before the eruption at 12:12:02 UT whereas the solid lines are from
12:15:02 UT.}
\end{figure*} 

\begin{figure*}\centering
\includegraphics[width=150mm]{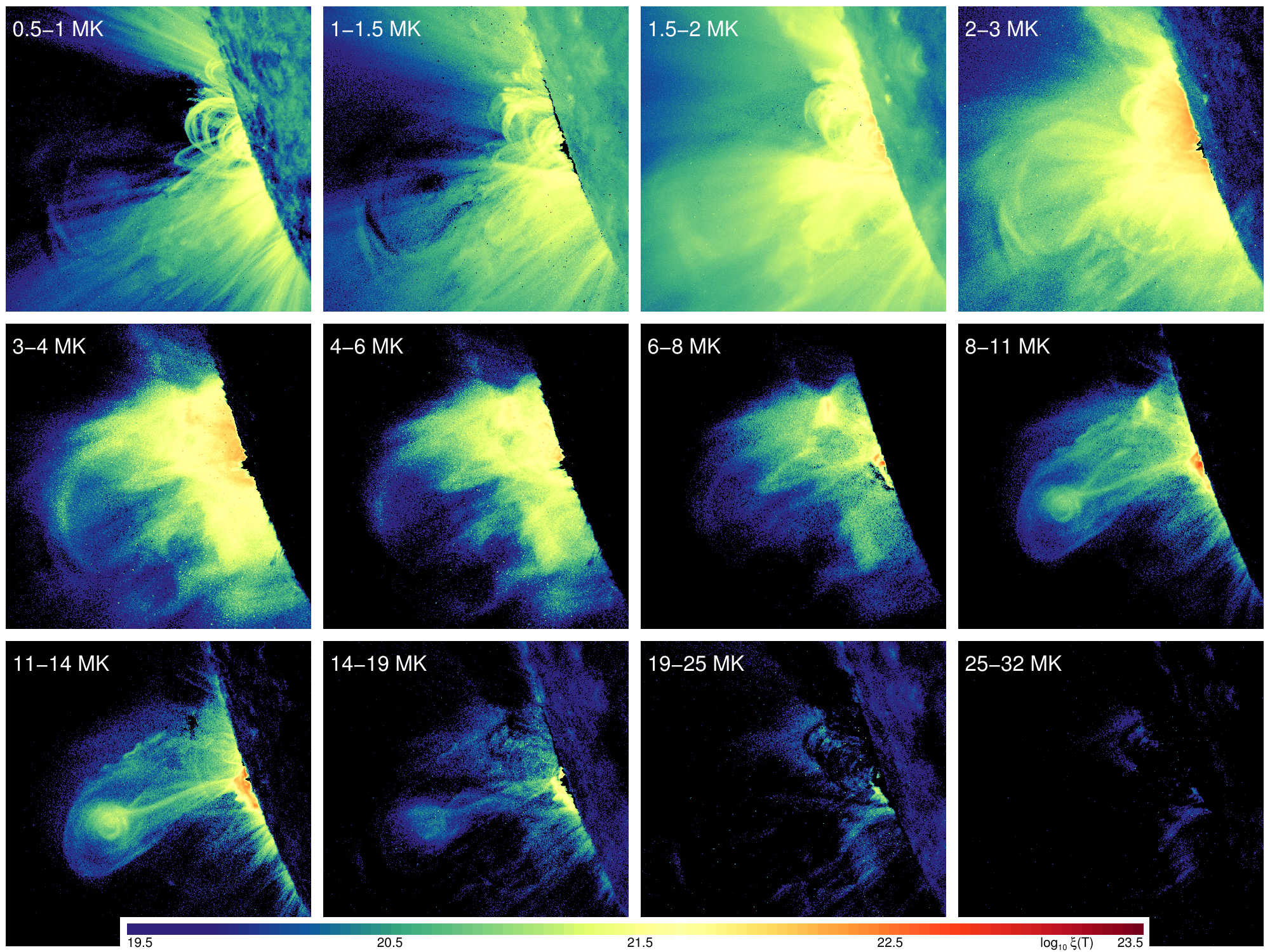}
\caption{\label{fig:DEM121502}DEM maps in different temperature ranges for the
3-Nov-2010 flare at 12:15:02 UT. The same pixel position in the twelve maps
corresponds to the regularized DEM solution $\xi(T)$ (as shown in detail for
eight pixel positions in Figure \ref{fig:DR121502}) to the set of data values
from the corresponding pixel in the six images shown
in Figure \ref{fig:aia121502}. The colour scale is $\log{\xi(T)}$ and is in
units of cm$^{-5}$K$^{-1}$.}
\end{figure*}

\section{DEM Maps of 03-Nov-2010 Eruptive Event}\label{sec:demres}

\subsection{Data Reduction}

The data between 12:12:02 UT and 12:19:02 UT (with a set of six  different
channel images every 12~seconds) were prepared to level 1.6 by deconvolving the
point spread function using \texttt{aia\_deconvolve\_richardsonlucy.pro} before
being processed with \texttt{aia\_prep.pro}. The co-alignment of the images were
checked using \texttt{aia\_coalign\_test.pro} and found to be sub-pixel
accurate. The resulting images were reduced down to sub-maps covering
-1150\arcsec to -850\arcsec in the x direction and and -500\arcsec to
-200\arcsec in the y direction, 501 by 501 pixels.

For each pixel position $(x,y)$ across the six different filters
($i=1\ldots6$) in the 12 second set $t$ provides the signal $g(x,y,t,i)$, in
units of DN s$^{-1}$px$^{-1}$, from which the associated errors $\delta
g(x,y,t,i)$ are calculated using the readout noise and photon counting
statistics. The set of six values, and their associated errors, with the
temperature response functions are passed to the regularized DEM map code, as
detailed in \S\ref{sec:demmeth}.
The SDO/AIA temperature response functions are calculated using both the
CHIANTI fix (an empirical correction for missing emission in the CHIANTI v7.0
database in channels 94\AA~ and 131\AA) and EVE normalisation (to match EVE
spectroscopic full-disk observations), further details available
online.\footnote{SSW/sdo/aia/response/chiantifix\_notes.txt}.
We use fairly broad temperature binning, twelve pseudo-logarithmic bins between
0.5 to 32MK, as to avoid a large number of temperature bins slowing down the
code. These are larger than the designed achievable temperature resolution
\citep{2010ApJ...708.1238J}, yet narrow enough to recover test models , from
testing the recovery of Gaussian model DEMs, as in \citet{2012A&A...539A.146H}.

\subsection{DEM from individual pixels}\label{sec:dempx}

The resulting DEMs found for eight distinct pixel locations in the images from
Figure \ref{fig:aia121502} are shown in Figure \ref{fig:DR121502}. These pixels
are shown for the 131\AA~and 211\AA~images but the data from all six
coronal-filter images were used to recover the DEMs. The vertical and
horizontal errors shown are calculated as detailed in \S\ref{sec:demregmet}. The
vertical error is just the linear propagation of the uncertainty on the source
data but the horizontal error is more complicated as it gives a measure of the
temperature spread (from the resolution matrix) as well as the quality of the
regularized solution (i.e. strong off-diagonal terms) in each bin, see
\citet{2012A&A...539A.146H} for further details. The DEMs of the different
positions show a variety of forms but clearly none are isothermal and the
majority cannot be represented by a single Gaussian model. Shown for
comparison are the DEMs obtained in these pixels before the eruption at
12:12:02 UT. 

For each specific pixel we have:
\begin{description}
\item[\textbf{Pixel 1 - Core of the erupting plasmoid:}] There are clearly two
distinct components, one peaking in the range 1.5-3MK, the other at
8-14MK.  The lower temperature one is likely the background corona along the
line-of-sight whereas the hotter from the plasmoid itself. The
pre-eruption DEM has no high temperature component;
\item[\textbf{Pixel 2 - Filament/stem behind the plasmoid:}] Again two
components at similar temperature ranges to the core of the plasmoid but the low temperature
component has larger emission (due to higher densities lower in
the corona compared to Pixel 1) and the higher temperature one has considerably
weaker emission (due to a smaller amount of heated material accumulating in this
location compared to the core, Pixel 1). The pre-eruption DEM
has no high temperature component;
\item[\textbf{Pixel 3 - High corona away from the event:}] A single component
covering 1-3MK with expected weak emission, since lower density higher in the
corona, and identical (within the errors bars) to the pre-eruption DEM;
\item[\textbf{Pixel 4 - Corona away from the event:}] A broader DEM with
slightly higher peak emission than Pixel 3, likely due to it being lower in the corona
but again peaking over 1-3MK, and similar (within the errors bars) to
the pre-eruption DEM;
\item[\textbf{Pixel 5 - Corona near the event:}] Higher emission and broader DEM
than similar location of Pixel 4 indicating the presence of more hot coronal
loops. The emission at high temperatures (above 6MK) is zero within the
errors bars. The pre-eruption DEM is slightly higher, though
consistent within the errors, so may indicate a slightly hotter and denser loop(s) moving out of
this region;
\item[\textbf{Pixel 6 - Low corona flare emission:}] Very high emission across
all temperatures with a bright peak at 1.5-3MK and 8-14MK for the flare
heated material expanding up above the limb to be visible. The
pre-eruption DEM shows emission across 6-8MK and 8-11MK but weaker 
at higher temperatures, which
might be due to the behind-the-limb flare;
\item[\textbf{Pixel 7 - Envelope just ahead of the plasmoid:}] Weaker emission
than the core of the plasmoid (Pixel 1) but again a similar DEM with two components,
peaking at 1.5-3MK and 8-14MK, the latter likely due to the accumulation
of hotter coronal emission in this leading front of the eruption. The pre-eruption
DEM does not show this hotter (8-14MK) emission;
\item[\textbf{Pixel 8 - Further ahead of the plasmoid:}] Similar to the low
temperature component of Pixel 7 (which is physically closer to the core of
the erupting material) but no higher temperature component. Consistent
(within the error bars) to the pre-eruption DEM.
\end{description}

\noindent Overall the DEMs in Figure
\ref{fig:DR121502} present a similar scenario to that found by previous authors 
\citep[]{2011ApJ...732L..25C,2012ApJ...750...44B}
who were studying the behaviour just in the different SDO/AIA channels, as
well as the DEM analysis from \citet{2012ApJ...761...62C}.
Given the inherent difficulties in recovering DEMs this is reassuring that their bulk
features are consistent with the original SDO/AIA data. Here we have a hot (8-11,
11-14MK) plasmoid core, well-defined with a filamentary stem-like structure
behind it, similarly hot. This connects down to the flare loops filling up with
even hotter material (14-19MK), which have expanded just above the limb,
consistent with the classic flux rope eruption model. At cooler temperatures
there is strong emission ahead of the erupting plasmoid but there is also a
line-of-sight coronal component (peaking about 1.5-2MK) in all the pixels
chosen.

\subsection{DEM Maps}\label{sec:demmaps}

\begin{figure*}\centering
\includegraphics[width=160mm]{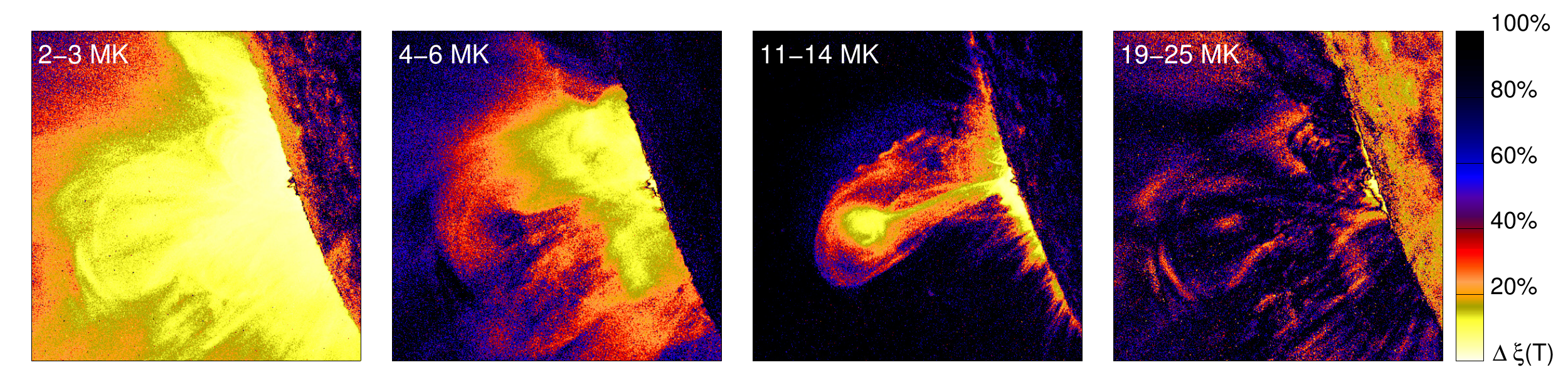}\\
\includegraphics[width=160mm]{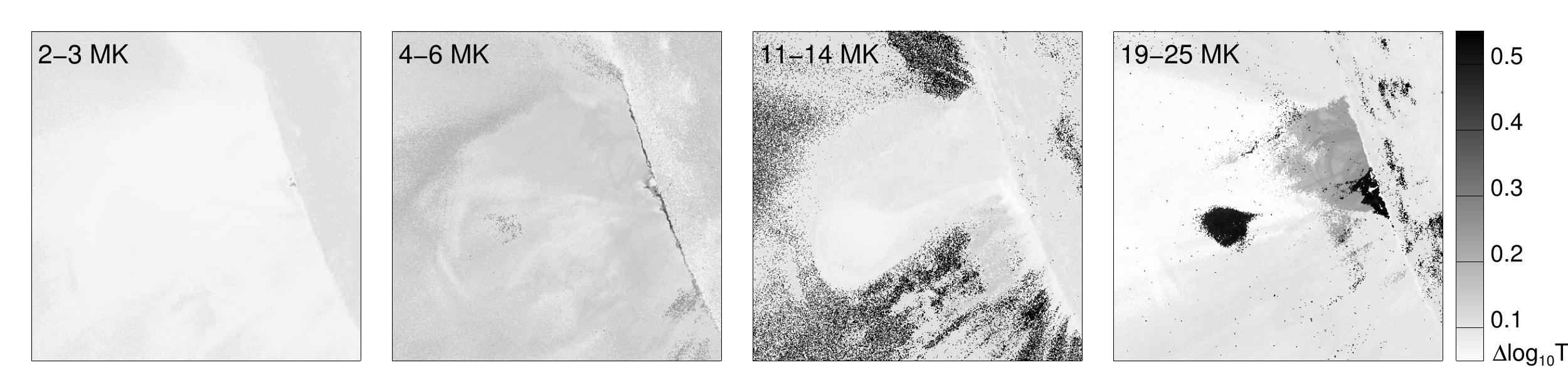}\\
\caption{\label{fig:errors}Maps of the (top row) vertical uncertainty
$\Delta \xi(T)$ and (bottom row) horizontal uncertainty, or temperature resolution
$\Delta \log{T}$, in the regularized DEM solution shown in Figure
\ref{fig:DEM121502} for four temperature ranges.}
\end{figure*}

To get a better understanding of the spatial structure of the event we
can create DEM maps for each temperature range, shown in Figure
\ref{fig:DEM121502}. Many pixel locations in these maps show emission across a
broad range of temperatures, highlighting the complex, multi-thermal nature of
the line-of-sight DEMs. The temperature maps for 0.5-1MK and 8-11MK do resemble
the images from the 171\AA~and 131\AA~channels respectively showing that the
emission here is mostly from one component of the channels' response. However
for the other temperature maps, features start to appear that are not apparent
in the SDO/AIA images themselves, such as there being little 4-6MK emission about
the core of the erupting plasmoid. This emphasizes the complicated temperature
response of SDO/AIA filters and the necessity of the DEM inference to interpret the
temperature or density structures in the corona. Before discussing these
spatial properties it is prudent to investigate which of the regularized DEMs
are robust solutions.

\begin{figure*}\centering \includegraphics[width=140mm]{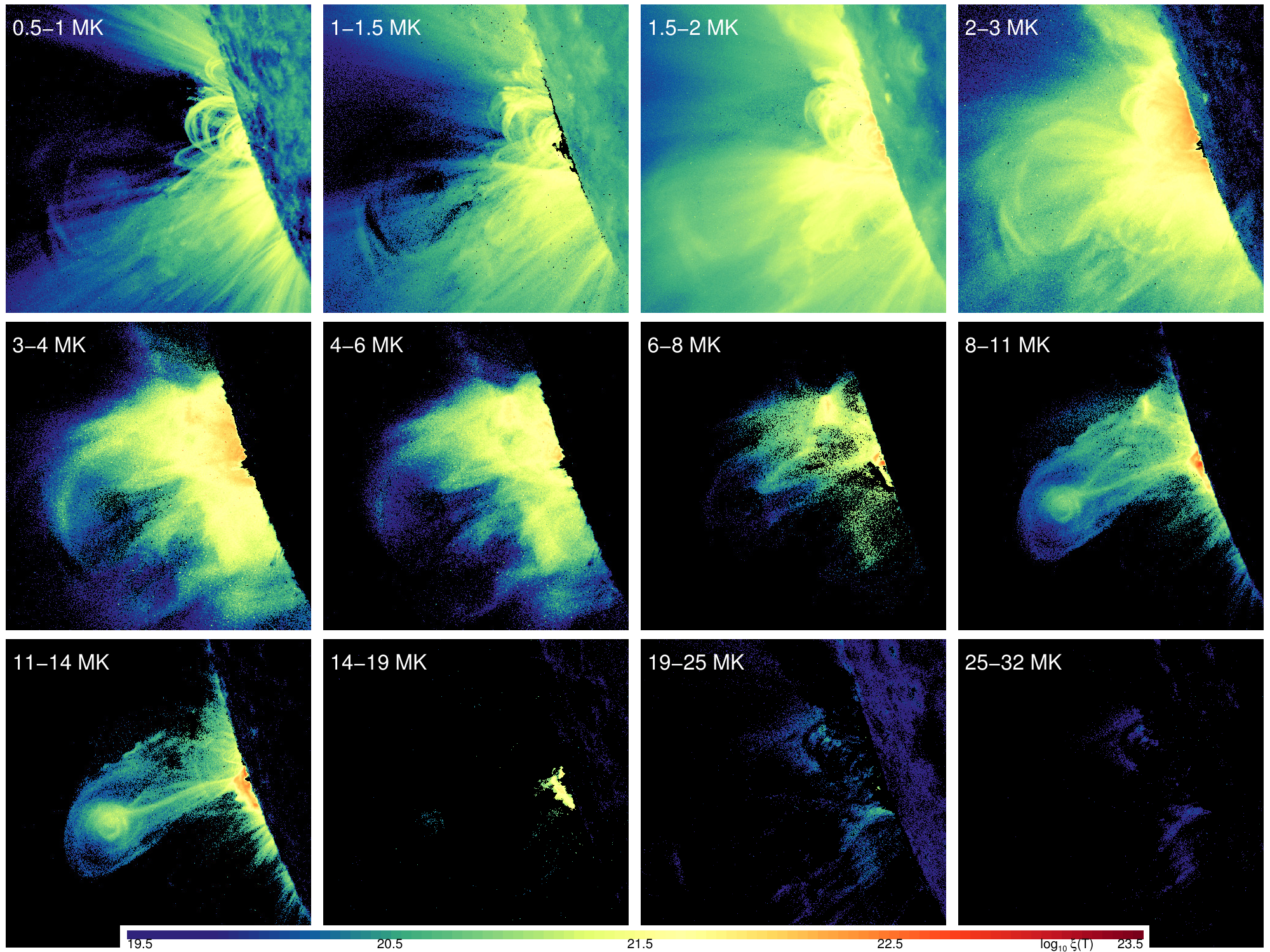}
\caption{\label{fig:gdDEM121502}DEM maps in different temperature ranges for the
3-Nov-2010 flare at 12:15:02 as shown in Figure \ref{fig:DEM121502} but only
showing pixels where $\Delta \xi(T) \le 30\%$ and $\Delta \log{T} \le 0.25$.
The colour scale is $\log{\xi(T)}$ and is in units of cm$^{-5}$K$^{-1}$.}
\end{figure*}

One major advantage of our regularization approach is that it produces
information about the vertical $\Delta \xi(T)$ and horizontal $\Delta
\log{T}$ uncertainty in the regularized solution (see \S\ref{sec:demregmet})
and so we can plot uncertainty maps, shown in Figure \ref{fig:errors}. The
vertical error $\Delta \xi(T)$ (top row, given as a percentage error of the DEM
signal) show the largest uncertainties are predominantly from the regions where
the emission measure is very low as well as at the highest temperatures. The
latter is due to a combination of SDO/AIA having lower sensitivity to the
temperatures $>20$MK and there also being little plasma emitting in this
temperature range. Hence the data, and also DEM, are very noisy. Conversely, at
the temperatures where SDO/AIA has a very strong response (about 0.5 to 3MK) and
there is bright solar emission, the signal, and hence DEM, has a very small
associated vertical error.

\begin{figure*}\centering
\includegraphics[width=180mm]{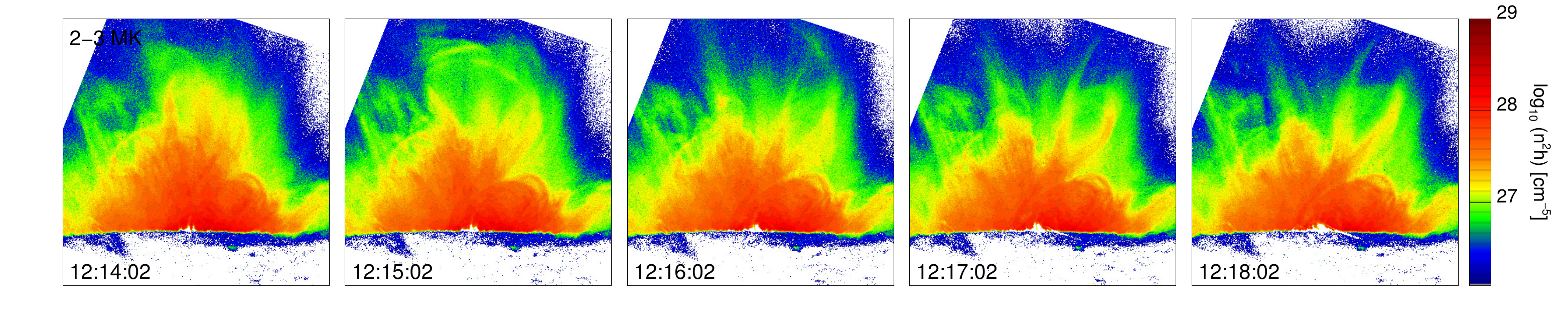}\\
\includegraphics[width=180mm]{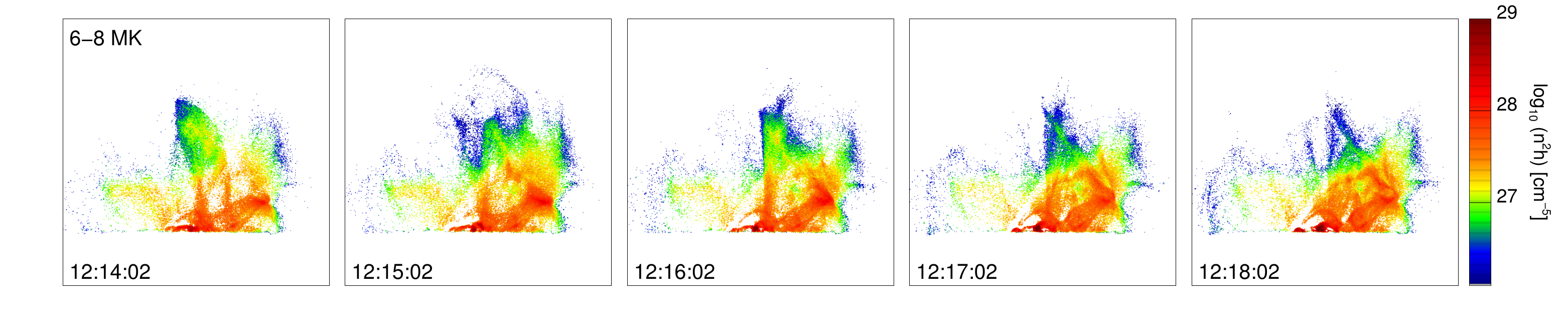}\\
\includegraphics[width=180mm]{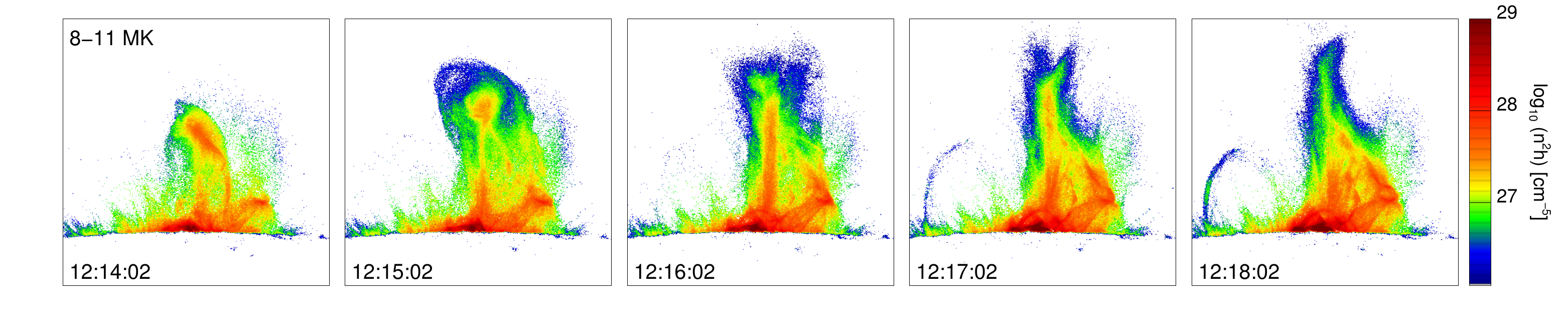}\\
\caption{\label{fig:ems} Emission Measure maps for various times in 2-3MK, 6-8MK
and 8-11MK (top, middle, bottom rows).
These maps have been rotated by $110^\circ$. The second column is of the time
interval shown in previous Figures. A movie of this is available with the
electronic version of this paper, covering the 1-1.5 to 14-19MK from
12:12:02 UT to 12:19:52 UT, see Figure \ref{fig:gdf}.}
\end{figure*}

The horizontal error in the solution (bottom row Figure
\ref{fig:errors}) are mostly slightly larger than the SDO/AIA minimum
theoretical  limit of temperature resolution $\Delta \log{T}=0.1$
\citep{2010ApJ...708.1238J}. The largest uncertainties again come from regions
where there is a weak signal and/or noisy data, this is particularly evident for
the largest temperatures.
But there are are also positions where the data is noisy in the hottest
temperature bands yet the horizontal error is small. This is due to the
regularization approach robustly determining that there is very weak/no emission
at these positions and temperature. Where the horizontal errors (temperature
resolution) are small it means that the emission is confined to that temperature
range but with large errors ($\geq 0.3$) it suggests the emission shown in that
temperature band could actually be attributed to a neighbouring temperature
bins. Therefore at temperatures where the horizontal uncertainty are large the
DEM solution should be treated with caution \citep[see][]{2012A&A...539A.146H}
and additional analysis of the residuals could be useful
\citep[e.g.][]{2003ApJ...595L.127P}.

\begin{figure}\centering \includegraphics[width=85mm]{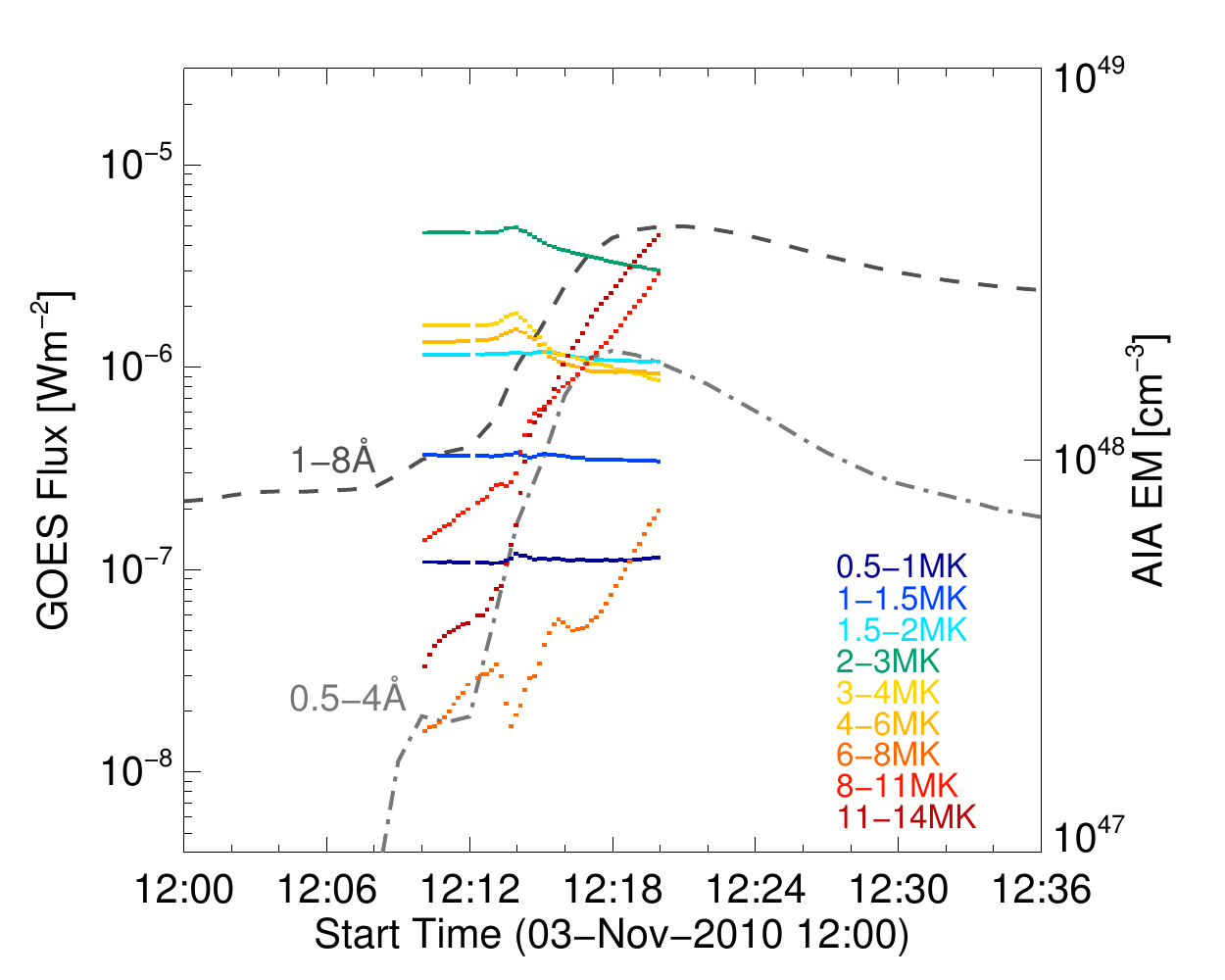}
\caption{\label{fig:goes}GOES SXR lightcurve of the 3-Nov-2010 flare C4.9 (grey
lines) with the spatially integrated emission measures derived from the SDO/AIA
observations for the chosen temperature ranges (colored lines).}
\end{figure}

Using this error information we can restrict the DEM maps to only show the
values with the smallest vertical and horizontal errors and this is shown for
$\Delta \xi(T) \le 30\%$ and $\Delta \log{T} \le 0.25$ in Figure
\ref{fig:gdDEM121502}. At the hottest temperature (25-32MK) there is virtually
no emission due to the largest uncertainties and poor confidence in the DEM
solution and any weak emission shown is likely unreliable. In 14-19MK
the emission is predominantly coming from the flare heated material that has
risen above the limb though there is a hint of very faint emission from the
plasmoid core. These loops show the strongest emission in 8-11 and
11-14MK but also in these temperature ranges is the plasmoid core, an envelope
of surrounding material and the filamentary stem behind it, all sharply visible.
The plasmoid core has slightly stronger emission in 11-14MK than 8-11MK. Only a
single thin stem structure is clearly defined in 11-14MK but two, in a V-shape
diverging away from the plasmoid core, are visible in 8-11MK. In both cases
their high temperatures, and the inflow region to these stems,
is not explained by the classical flux rope eruption model. Some additional
heating maybe present and the Kelvin-Helmholtz waves to the top of the erupting
material are a signature of a reconnection layer surrounding the erupting
flux rope, producing a secondary means of releasing energy
\citep{2011ApJ...729L...8F,2012foullon}. 

\begin{figure*}\centering
\includegraphics[width=180mm]{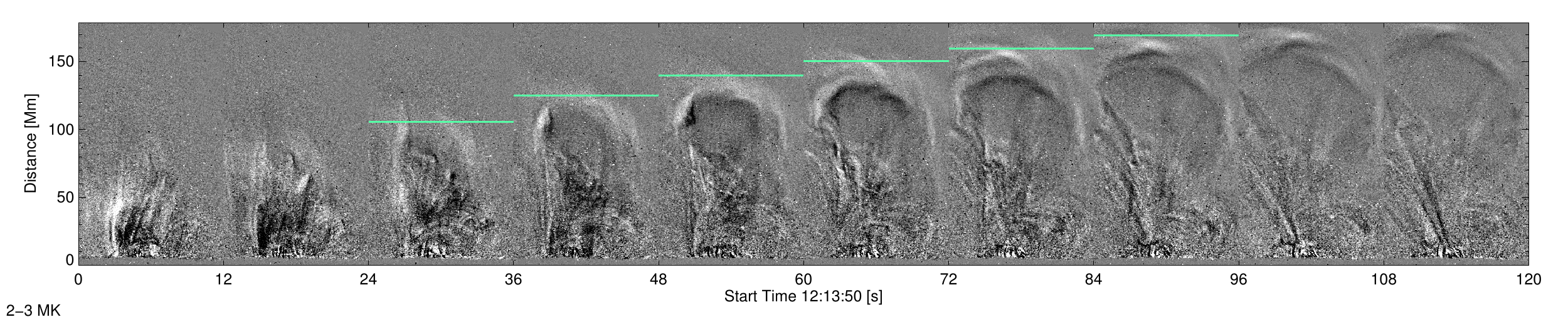}\\
\includegraphics[width=180mm]{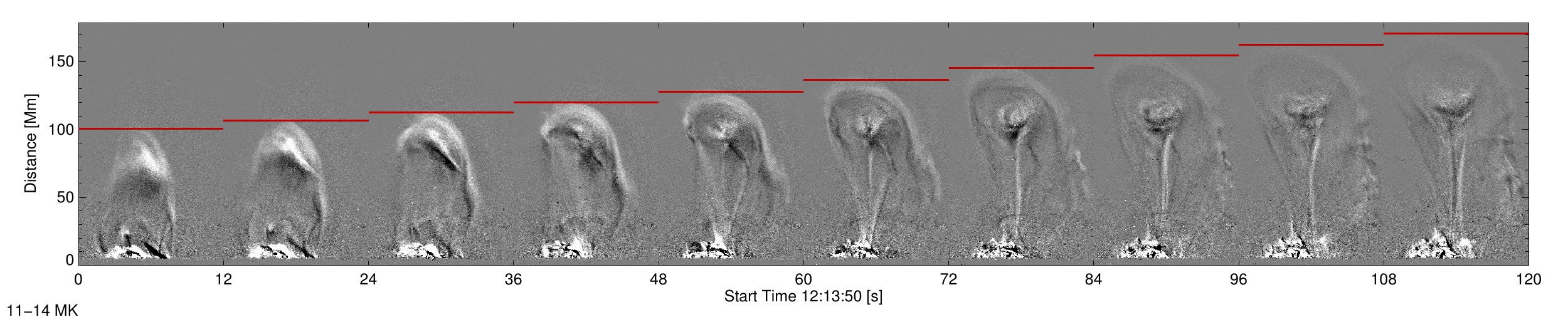}\\
\caption{\label{fig:rddem}
Running differences of the DEM maps
($\xi(T,x,y,t_i)-\xi(T,x,y,t_{i-1})$) for 2-3MK (top row) and 11-14MK (bottom
row). The leading edge of the erupting plasma was tracked and is indicated by
the red horizontal lines The resulting height profiles as a function of time for
these and other temperature ranges is shown in Figure \ref{fig:tvh} and
\ref{fig:tvv}. These maps have been rotated by $110^\circ$.}
\end{figure*}

At cooler temperatures ($\le$6MK) the core of the plasmoid disappears, which is
likely the cause of the dimming observed in some SDO/AIA images but it is
not clear whether this is due to heating or mass loss. To try and understand
this we look at the time evolution of the maps in \S\ref{sec:demtime}. The
V-shaped structure of the stems is still present though, clearly in 6-8MK but
slightly obscured by the bright surrounding coronal emission in 3-4 and 4-6MK.
Ahead of the plasmoid core we start to see the leading edge of a front of
material, visible in 2-3, 3-4 and 4-6MK. This density increase of material is
likely due to a compression front being driven ahead of the erupting plasmoid.
At the coolest temperatures (0.5-1, 1-1.5 and 1.5-2MK) there is some faint
evidence of the leading edge moving out but it is dominated by the emission from
several hot coronal loops associated with the occulted active region and flare.

\subsection{Time Evolution}\label{sec:demtime}

As the 03-Nov-2010 solar eruption is highly dynamic we investigate the DEM maps
as a function of time, which we have shown in three temperature ranges at 1
minute intervals in Figure \ref{fig:ems}.
A movie of this is available with the electronic version of this paper, covering
1-1.5 to 14-19MK from 12:12:02 UT to 12:19:52 UT, see Figure
\ref{fig:gdf}. We have also converted the DEM maps into EM maps by removing the
factor $\ln{10} T\mathrm{d}\log{T}$, resulting in the values shown being
$EM=n^2 h$ [cm$^{-5}$], making it easier to calculate densities (see
\S\ref{sec:deneng}). Again we have only shown the emission in the pixels and
temperature ranges that have $\Delta \xi(T) \le 30\%$ and $\Delta \log{T} \le
0.25$, as in Figure \ref{fig:gdDEM121502}.

\begin{figure}\centering \includegraphics[width=85mm]{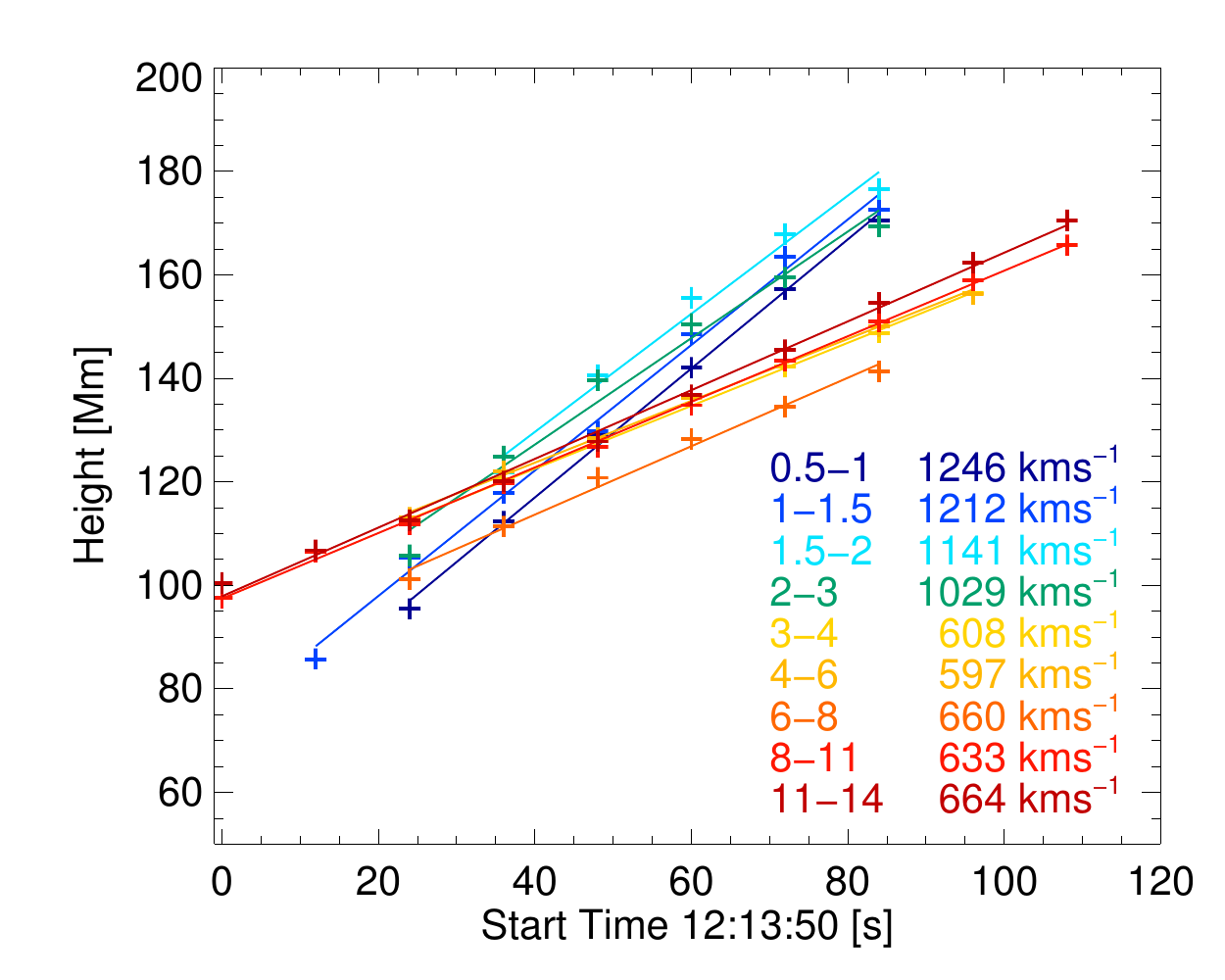}
\caption{\label{fig:tvh}
Height (above an arbitrary origin) versus time profiles of the leading
edge of the erupting plasma for each of the temperature ranges the propagation
could be reliably tracked.}
\end{figure}

In the 2-3MK temperature range (top row Figure \ref{fig:ems}) we see the leading
edge of the eruption clearly at 12:15:02 UT but afterwards gone and the loops
bending/collapsing back. There is some strong emission ahead of erupting flux
rope at 12:14:02 UT but that disappears at later times, likely erupted. Lower in
the corona the loops barely change over time and only the highest ones modify,
being bent back after the eruption, leaving a void where the stem forms in
higher temperatures. In 6-8MK (middle row Figure \ref{fig:ems}) we see the
initial eruption of the plasmoid core and stems behind it at 12:14:02 UT but at
later times there is a void of material at this temperature, again suggestive of
material loss (or spreading out) due to the eruption mass loss.
There is increased emission from the loops that have bundled up below the
eruption to the right indicating possible heating of material in these
contracting loops.

The erupting plasmoid and the filamentary stem structure connecting it
to the lower solar atmosphere is sharply defined in 8-11MK (bottom row Figure
\ref{fig:ems}). The strong emission from the core of the plasmoid quickly
weakens as it moves higher leaving the strongest emission from the vertical
stem behind it. It would be tempting to immediately classify this structure as a
current sheet, given the superficial similarity to the classical eruption
model. However this stem is relatively thick and appears to be
made up of several thinner structures. This maybe a projection effect with
multiple thin stem misaligned along the line-of-sight. A bigger issue is that
this structure shows strong emission in this hot (11-14MK) temperature range and
again the classical model is unable to explain the presence of such hot material
inflowing to the current sheet. Below to the right of the eruption (or to the
top right in the original orientation) hot loops bundle up between 12:16:02 UT to
12:18:02 UT. This is associated with Kelvin-Helmholtz waves
\citep{2011ApJ...729L...8F} and possible density increases due to outflowing
material \citep{2012ApJ...754...13S}. In the classic eruptive model this region
would be associated with shrinking loops and the outflow from the reconnection
region. However the dynamics here are clearly more complicated and difficult to
unambiguously ascertain due to the strong emission lower in the atmosphere. In
the 12:17:02 UT and 12:18:02 UT images another large coronal loop develops to the
left of the main eruption. This fainter emission in 11-14MK is likely flare
heated material that has expanded into this large loop.
STEREO-B/SECCHI images from behind the limb do show a separate core flaring
region located at the base of this hot loop \citep{2012ApJ...754...13S}.

By summing over the whole of the region shown in Figure \ref{fig:ems} we can
produce a time profile in each temperature band, shown in Figure \ref{fig:goes}
in comparison to the GOES SXR emission. We have computed this between 12:12:02 UT
and 12:19:50UT (the time range shown in the movie available in the electronic
version of this paper, see Figure \ref{fig:gdf}) as saturation from the
flaring loops starts to dominate and produces unreliable EM maps in these
regions. From Figure \ref{fig:goes} we see that there is a slight
increase in most of the temperature ranges as the eruption begins (12:13UT)
which then drops off as the material is ejected (12:15UT), suggesting
this initial change is due to the heating, then the loss of coronal plasma.
After this the hottest temperature ranges (6-8, 8-11 and 11-14MK) increase in a similar
manner (though slightly delayed) to the SXR emission observed by GOES.
This is initially due to further heating in the current sheet and the shrinkage
of loops below the eruption with heated material outflowing the reconnection
region. The later sharp increase is likely due to the bright post-flare
loops rising above the limb. At the cooler temperatures it will be later in the
event before the current sheet and flare heated material (if visible above the
limb) cool into these temperature range. This explains the decrease in the
emission at these temperatures after the loss of material during the eruption.
The very small decrease at the smallest temperatures (0.5-1, 1-1.5,
1.5-2MK) suggest that this emission is dominated by the background and
line-of-sight coronal emission over the whole region.

\begin{figure}\centering
\includegraphics[width=85mm]{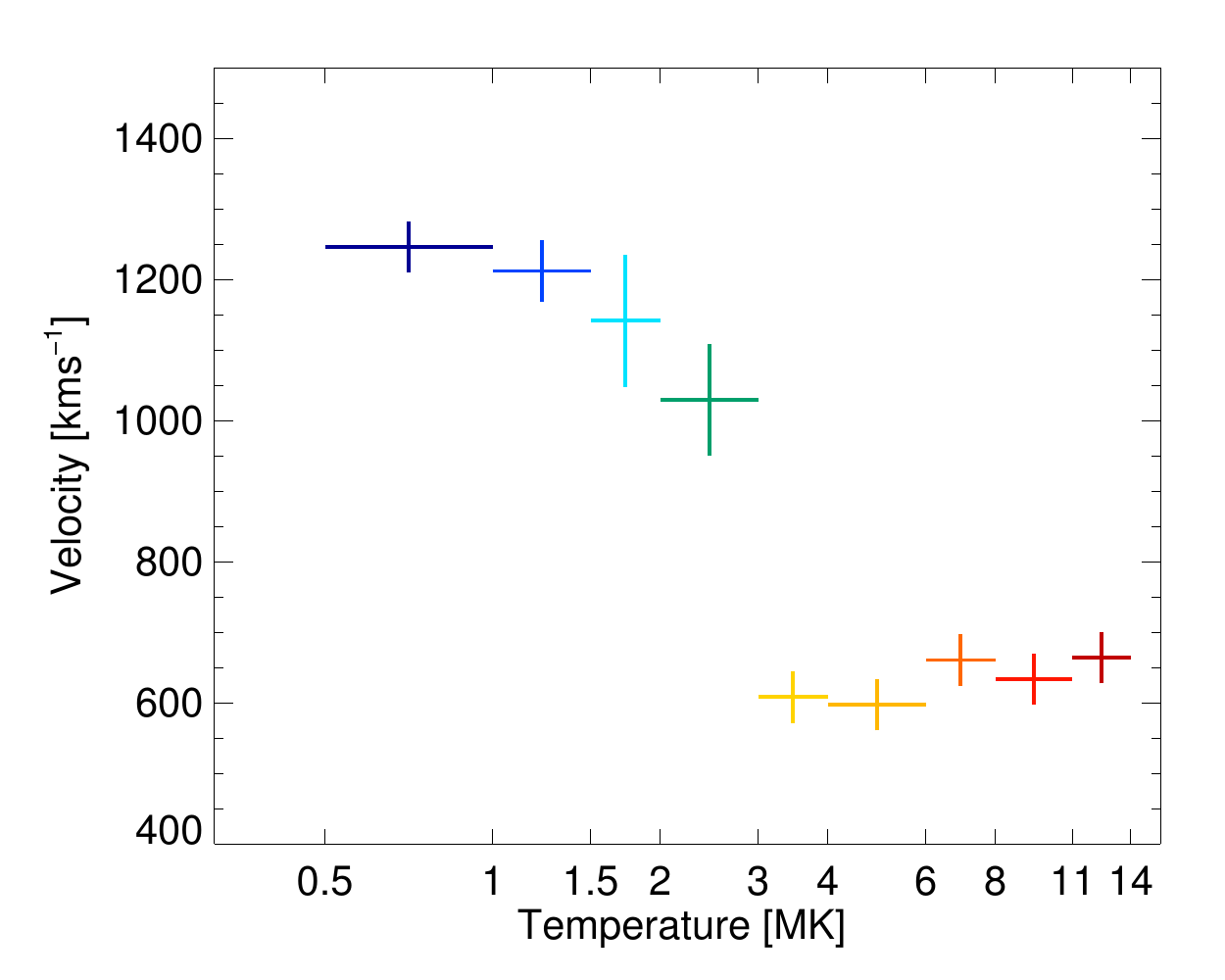}
\caption{\label{fig:tvv}Velocity of the leading edge of the eruption as
a function of temperature, obtained from linear fits to the data shown in Figure
\ref{fig:tvh}. The vertical uncertainty is taken as the largest of the error
in the parameter from the linear fits shown in Figure \ref{fig:tvh} and three
AIA pixels travelled in 12s.}
\end{figure}

\begin{figure}\centering
\includegraphics[width=85mm]{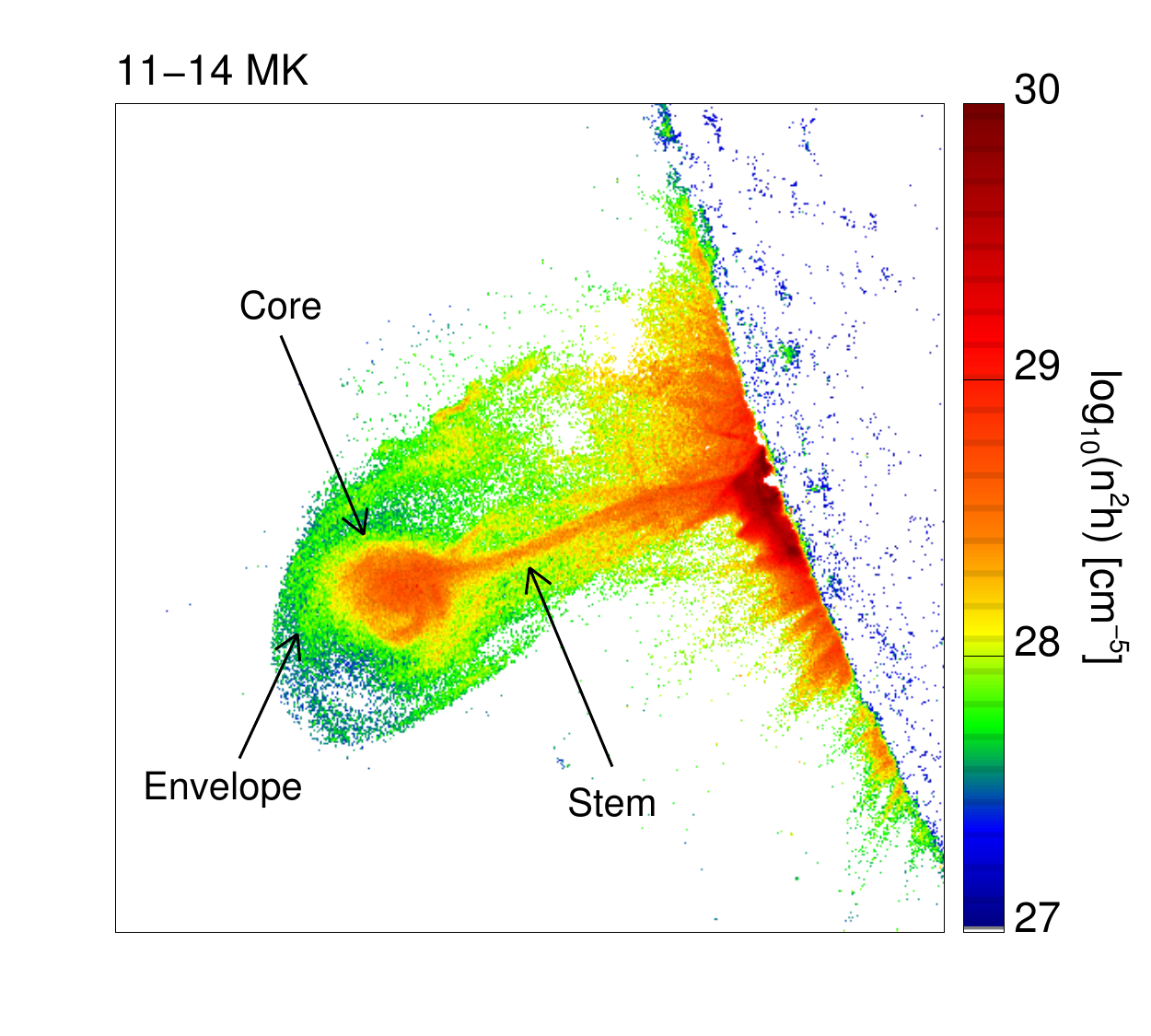}
\caption{\label{fig:emlab}
Emission Measure map for 11-14MK at 12:15:02 UT with the different components
identified.}
\end{figure}

\begin{figure*}\centering
\includegraphics[width=180mm]{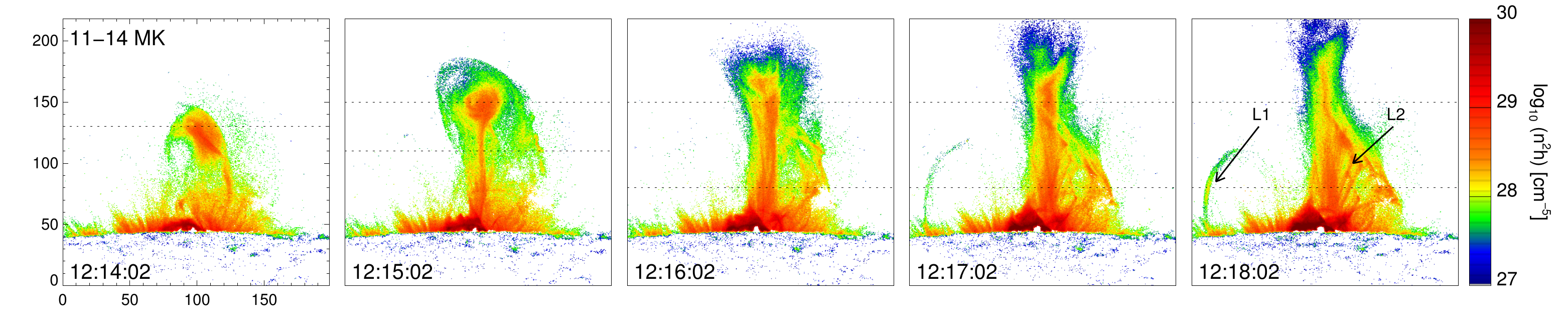}\\
\includegraphics[width=180mm]{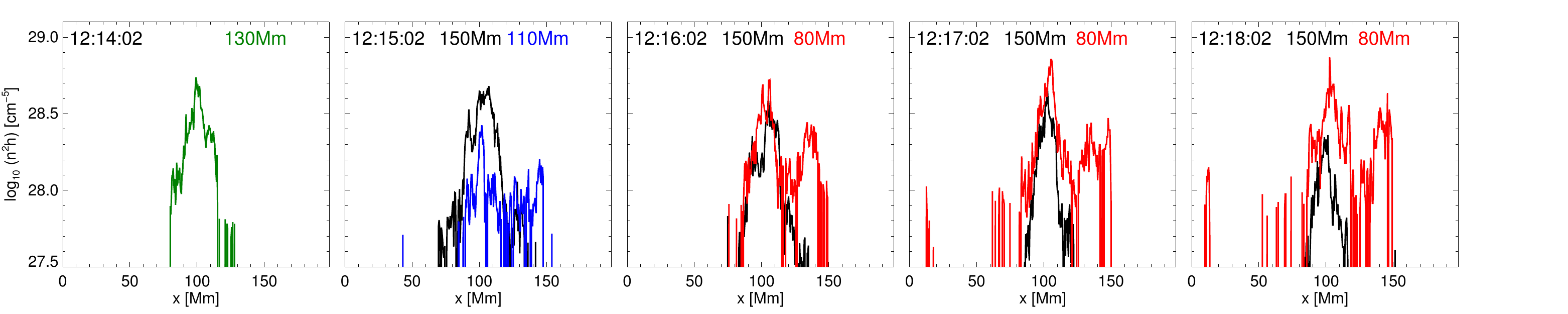}\\
\caption{\label{fig:emsengs}(Top) Emission Measure maps for various times in
11-14 MK. (Bottom) The emission measure profile at various heights for one
minute time intervals in 11-14 MK, indicated by the dotted lines in the top
panel.}
\end{figure*}

To investigate the time profile of the erupting material in different
temperature ranges we have manually tracked the front of the emission in running
difference maps of the emission measure, i.e.
$\xi(T,x,y,t_i)-\xi(T,x,y,t_{i-1})$. This is shown in Figure \ref{fig:rddem} for
2-3 and 11-14MK, with black indicating a decrease in emission, white an increase
and the colour lines showing the manually tracked front of the eruption.
This type of analysis has been performed before but in individual SDO/AIA wavelength
channels \citep{2011ApJ...732L..25C,2012ApJ...750...44B} and not for specific
temperature ranges. The former has the advantage of a clearer time evolution as
the channels are a $\approx$3s duration image every 12s whereas the EM maps are
effectively 12s cadence due to the use of six different channels in their
calculation. This makes the determination of the leading edge tricky and the
resulting height (above an arbitrary position) as a function of time for the
different temperature ranges are shown in Figure \ref{fig:tvh}. Here we have
linearly fitted the data to produce a velocity as a function of
temperature. These velocities and there associated error (the largest
between the error in the fitted parameter and from three SDO/AIA pixels travelled in
12s) are shown in Figure \ref{fig:tvv}. In the cooler temperature range the
leading edge of the eruption travels at 1\,246 kms$^{-1}$ in 0.5-1MK decreasing
to 1\,029 kms$^{-1}$ in 2-3MK. This general trend of a slower eruption at higher
temperature is more evident above 3-4MK once we are in the range associated with
the plasmoid eruption itself. From 3-4Mk to 11-14MK the velocities are
consistently between 597 kms$^{-1}$ to 664 kms$^{-1}$ suggesting the
multi-thermal structure is moving outwards together. These two distinct
structures at different velocities are consistent with those found by tracking
the eruption in the SDO/AIA channels
\citep{2011ApJ...732L..25C,2012ApJ...750...44B}, with the 131\AA~ and 335\AA~ at
667 and 747 kms$^{-1}$ and the cooler 211\AA~and 193\AA~at 1\,154 and 1\,439
kms$^{-1}$ \citep{2012ApJ...750...44B}.

\begin{table*}
\sidecaption
\caption{Thermal parameters for the different features shown in Figures
\ref{fig:emlab} and \ref{fig:emsengs}.}
\label{table:params}
\centering
\begin{tabular}{ccccc|cccc|ccc}
\hline
\hline
&&&&&(h=w)&&\multicolumn{2}{c|}{Energy}&
(h=100Mm)&\multicolumn{2}{c}{Energy}\\
Time & Feature & EM & Width & Length   &
Density  &Vol& Thermal   & Kinetic
&Density  & Thermal   & Kinetic\\
&  & $10^{28}$cm$^{-5}$ & Mm & Mm   &
$10^{9}$cm$^{-3}$ &$10^{28}$cm$^{3}$ &
$10^{29}$erg & $10^{29}$erg&$10^{9}$cm$^{-3}$ &
$10^{29}$erg & $10^{29}$erg\\
\hline
12:14:02 & Core & $2.5$ & 25 & 40 &
$3.2$ & 2.5&$4.1$ &2.9&1.6&8.2&5.8\\
\hline
12:15:02 & Core & $3.9$ & 30 & 30 &
$3.6$ &2.7&$5.1$  &3.6&2.0&9.3&6.5\\
& Envelope & $0.5$ & 65 & 100 &
$0.9$ &39.6&$18.0$  &12.6&0.7&23.8&16.7\\
& Stem & $2.5$ & 7 & 80 &
$6.0$ &0.4&$1.2$  &&1.6&4.6&\\
\hline
12:16:02 & Stem & $2.0$ & 25 & 100 &
$2.8$ &6.3&$9.1$  &&1.4&18.3&\\
\hline
12:17:02 & Stem & $3.2$ & 20 & 110 &
$4.0$ &4.4&$9.1$  &&1.8&20.3&\\
\hline
12:18:02 & Stem & $1.6$ & 10 & 130 &
$4.0$ &1.3& $2.7$ &&1.3&8.5&\\
\hline
\end{tabular}
\end{table*}

\subsection{Density and Thermal Energy}\label{sec:deneng}

To recover the density from the EM maps an estimate of the line-of-sight
depth $h$ of the emission is required. As the structures are sharply defined at
higher temperatures (visbile in Figure \ref{fig:emlab}, labelling the three
distinct components: plasmoid core, envelope and stem) we can use their apparent
width in the EM maps as the depth assuming spherical/cylindrical symmetry. We
focus just on 11-14MK here as the structures are similar in shape to those in
8-11MK (bottom row of Figure \ref{fig:ems}) and at lower temperatures this would
be difficult as the shape of the structures is far harder to estimate reliably.
The density of the sheath of the Kelvin-Helmholtz waves has been estimated at
lower temperatures (about 4MK) using emission measures found using both our
regularized approach and Gaussian model forward-fitting techniques, finding
similar values, and is the subject of a separate publication
\citep{2012foullon}.

The EM maps in 11-14MK for 5 time intervals during the plasmoid eruption are
shown in Figure \ref{fig:emsengs} (top row). Here we have plotted the EM as a
function of position (bottom row) for various slices through the erupting
structure, indicated in the EM maps by dotted horizontal lines. From these
slices we can determine a width of the structures of interest: the erupting
plasmoid core, envelope and stem.  Assuming a spherical/cylindrical geometry in
these structure we can use the width as the depth, giving the line-of-sight
extent $h$. Since $EM=n^2 h$ this readily gives an estimate of the density in
the structure and these values for the various structures at the different times
are given in Table \ref{table:params}. Here we see that the density in the core
region (about $3\times10^9$cm$^{-3}$) does not change between the two time
intervals shown before it fully erupts. The envelope of material
surrounding the plasmoid core has a lower density ($3\times10^8$cm$^{-3}$) but
still shows a sharp transition to the background corona as that has effectively
no emission in this temperature range.

The density in the stem region does decrease over time but that is initially due
to a widening of the structure (seen at 12:16:02 UT onwards) rather than a
reduction in the emission measure. As discussed before, it is difficult to
attribute such a wide hot structure to the monolithic current sheet as proposed
by the classic eruption model.  If we instead assuming the structures are more
elongated, using a line-of-sight component of $h=100$Mm (shown in Table
\ref{table:params}), we find that the density of the stem does not vary much
with time. In fact these slightly smaller densities (about $10^9$cm$^{-3}$) are
fairly similar for both the core and stem regions.

The thermal energy of this emission can be calculated via
$W_\mathrm{T}=3nVk_\mathrm{B} T$, where we take $T=12.5$MK the middle of the
temperature range. We estimate the volume using $V=whL$, with $w$ the width, $h$
the line-of-sight component and $L$ the extent perpendicular in the image to
$w$. This is found using a filling factor of unity, and as this could be
smaller, the values we calculate are upper limits to the thermal energy. Note
that for the envelope region we have calculated the volume of the region minus
the volume of the plasmoid core. The energies are shown in Table
\ref{table:params} for the times and regions in Figure \ref{fig:emsengs}. We
find that the thermal energy in this temperature range as the core erupts
increases slightly ($4.1$ to $5.1\times10^{29}$ erg) whereas the energy in the
stem structure greatly increases just after the impulsive eruption ($1.2$ to
$9.1\times10^{29}$ erg). This is due to the widening of the stem whilst
maintaining the same emission measure per pixel within the stem.
This weakens by 12:18:02 UT when the emission measure starts to drop and the stem
narrows. If we again assume the structures are elongated, with line-of-sight
$h=100$Mm, we get a smaller variation between the thermal energy of the
different structures and how they evolve with time, with energies of about
$10^{30}$ erg. In both cases there is an increase in the thermal energy of the
stem which is consistent with the idea of continuous energy release in this
region. The envelope of hot material around the plasmoid core has a considerably
higher thermal energy but that is because it is larger structure.

We can also calculate the kinetic energy for the structures which clearly
propagate outwards, the plasmoid envelope and core. The kinetic energy is taken
as $W_\mathrm{K}=nVm_\mathrm{p}v^2/2$ where $m_\mathrm{p}$ is the proton mass
and the velocity is $v=660$ kms$^{-1}$ from Figure \ref{fig:tvh}.
These are again given in Table \ref{table:params} for $h=w$ and $h=100$Mm. Here
we find that the thermal energy (in 11-14MK) is consistently slightly larger
than the kinetic energy of the material in this temperature range. A
higher thermal to kinetic energy was also found previously by
\citet{2010ApJ...711...75L}. This similarity between the thermal and kinetic
energies is typical for CMEs from observations higher in the corona
\citep[e.g.][]{2000JGR...10523153F}. Our values for the energies are also
smaller than those for a ``moderately'' large CME \citep{2000JGR...10523153F}
($10^{30}$ versus $10^{32}$ erg) which is expected as our energy is just from
one component of the CME and in one temperature range.

\subsubsection{Coronal Loop Density}

In the later time intervals in Figure \ref{fig:emsengs} (12:18:02 UT) there are
distinctive coronal loops associated with the flaring region (L1) and the
shrinkage below the eruption (L2). Again assuming $w=h$ we can calculate the
loop density, taking them to be about 2Mm, though we do not estimate the energy
as we cannot determine a reliable loop length.
The large coronal loop (L1), associated with the core of the occulted flaring
region, is visible from 12:17:02 UT onwards, and its density increases as more
hot material fills the loop (from $5.6\times10^9$cm$^{-3}$ to
$7.1\times10^9$cm$^{-3}$). The loops below the ejected material (L2) have an
emission measure that doubles between 12:17:02 UT and 12:18:02 UT, resulting in a
density increase to $1.3\times10^{10}$cm$^{-3}$ from $8.9\times10^{9}$cm$^{-3}$.
This increase in heated material maybe due to outflows from the reconnection
region in the stem structure \citep{2012ApJ...754...13S} or the reconnection
layer above the loops where Kelvin-Helmholtz waves were located
\citep{2012foullon}.

\section{Discussion \& Conclusions}\label{sec:disc}

Regularized inversion of multi-filter observations provides an essential tool to
study the temporal and spatial evolution of multi-thermal plasma during eruptive
events.  The regularized maps of an erupting CME presented in the paper allow
the basic plasma properties (integrated along the line-of-sight) to be inferred
and calculation of the density and energetics of various parts of the eruption.
We find that the leading edge of the eruption travels between 1\,029-1\,246 km
s$^{-1}$ and has the temperature range $\le$2-3MK followed by slower 597-664 km
s$^{-1}$  but hotter plasma $\ge$3-4MK. The erupting core and stem which could
be interpreted as a plasmoid with a current sheet appears sharply defined in
11-14MK. The width of the stem/current sheet we find (7-25Mm) is far
larger than that expected in the classical reconnection model, and somewhat
smaller than the predicted $<1Mm$ \citep{2009A&A...499..905V}.
This value is in fact smaller than those found higher in the corona
\citep[e.g.][]{2002ApJ...575.1116C, 2012ApJ...751...21L} suggesting continuous
expansions as the CME lifts up. The stem/current sheet width we determine is
however at a higher temperature range and made with a higher spatial resolution
observations than these previous results.

During the impulsive phase, we find that the density is about $3\times10^9$
cm$^{-3}$, $6\times10^9$ cm$^{-3}$, $9\times10^8$ cm$^{-3}$ in the plasmoid
core, stem and surrounding envelope of material. This gives thermal energy
estimates of $5\times10^{29}$ erg, $1\times10^{29}$ erg and $2\times10^{30}$
erg. These are slightly larger than the values found for the kinetic
energy of the erupting envelope and core, which is consistent with previous CME
energy estimations \citep{2010ApJ...711...75L}. The observations also show the
increase of the thermal energy of the core (plasmoid) and of the stem (current
sheet) as the CME is rising from 12:14:02 UT to 12:17:02 UT, suggesting ongoing
energy release. The increase of energy is more noticeable for the stem with it
more than doubling over 3 minutes.  This is consistent with the predictions of
recent reconnection models \citep[e.g.]{2010JGRA..11509206M,2010ApJ...721.1547R}
with the energy released in the current sheet pumped into the plasmoid.

The reconstructed DEM maps show the multi-thermal nature of the corona, which is
unsurprising given that we have a hot plasmoid erupting through the
cooler surrounding corona, two distinct structures with different temperature
characteristics. Our DEM maps show in greater detail the same basic physical
picture that was found by \citet{2012ApJ...761...62C} using a different DEM
reconstruction method for a few regions in the CME. Namely that the bright
leading front of the CME is composed of a corona-like temperature distribution,
with the increased emission due to the accumulation of material, whereas the
erupting core contains heated material ($>8$MK). All DEM solutions however
should be treated carefully given the uncertainties in the temperature response
functions and the possibly broad temperature resolution
\citep{2012ApJ...758...54T}. Used carefully, with these issues kept in mind,
regularized DEM maps present a useful tool for exploring the dynamics of heating
in the solar corona with SDO/AIA observations.

\begin{acknowledgements}

This work is supported by a STFC grant ST/I001808/1 (IGH,EPK). Financial support
by the European Commission through the FP7 HESPE network (FP7-2010-SPACE-263086)
is gratefully acknowledged (EPK).

\end{acknowledgements}


\Online
\begin{appendix}

\section{EM Movie of the Eruption}

\begin{figure}[h]\centering \includegraphics[width=85mm]{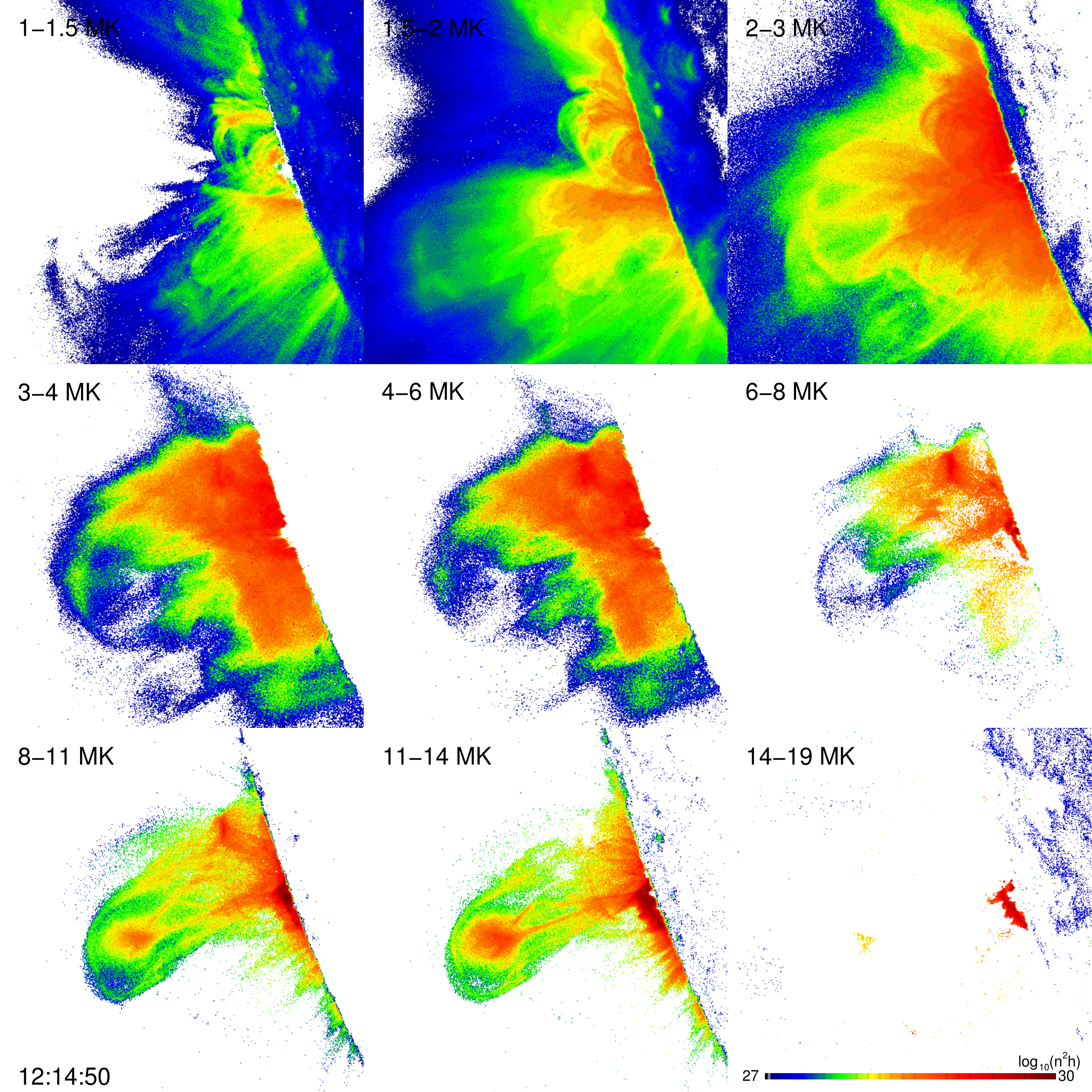}
\caption{\label{fig:gdf}One frame of the movie, available online, showing the EM
maps, $\log{n^2h}$ in units of cm$^{-5}$, for the temperature 
ranges 1-1.5 through to 14-19MK, starting at 12:12:02 UT running till
12:19:52 UT.}
\end{figure}

\end{appendix}

\end{document}